# Informational Size in School Choice

Di Feng[*] and Yun Liu[†]

July 15, 2024

**Abstract**

This paper introduces a novel measurement of *informational size* to school choice problems, which inherits its ideas from Mount and Reiter (1974). This concept measures a matching mechanism's information size by counting the maximal *relevant* preference and priority rankings to *secure* a certain pairwise assignment of a student to a school across all possible matching problems. Our analysis uncovers two key insights. First, the three prominent strategy-proof matching mechanisms, the deferred acceptance (DA) mechanism, the top trading cycles (TTC) mechanism, and the serial dictatorship (SD) mechanism, is *(strictly) less informative than* the non-strategy-proof immediate acceptance (IA) mechanism. This result highlights a previously omitted advantage of IA in term of its information demand, which partially explain the its popularity in real-world matching problems especially when acquiring information is both pecuniarily and cognitively costly. Second, when the matching problem contains at least four students, the TTC demands less information compared to the DA to implement a desired allocation. The issue of comparison between TTC and DA has puzzled researchers both in theory (Gonczarowski and Thomas, 2023) and in experiment (Hakimov and Kübler, 2021). Our result responds to this issue from an informational perspective: in experiments with relatively fewer students, agents tend to prefer DA over TTC as DA requires fewer information to secure one's allocation in all problems (Guillen and Veszteg, 2021), while the opposite is true when the market size increases (Pais et al., 2011). Among others, our informational size concept offers a new perspective to understand the differences in auditability (Grigoryan and Möller, 2024), manipulation vulnerability (Pathak and Sönmez, 2013), and privacy protection (Haupt and Hitzig, 2022), among some commonly used matching mechanisms.

---

[*]Corresponding author. Department of Finance, Dongbei University of Finance and Economics, Dalian 116025, China. Email: dfeng@dufe.edu.cn.
[†]Center for Economic Research, Shandong University, Jinan 250100, China. Email: yunliucer@sdu.edu.cn.

# 1 Introduction

When designing mechanisms for real-life matching markets, we usually assume that the benevolent social planner is able to acquire all students' private information to determine a desired allocation for students, even if certain pieces of information are not necessary.[1] Such complete information awareness premise simplifies analysis, it nevertheless overlooks the role of information on evaluating the performance of different mechanisms in various real-life scenarios. For instance, even though theoretical analysis indicate that participant's optimal strategy (weak dominant strategy) is truth-telling in strategy-proof mechanisms, many participants do not necessarily choose their optimal strategies and even misreport their preferences under strategy-proof mechanisms, when the information environments vary (see Hakimov and Kübler (2021) for an excellent survey). Also, students may have privacy concerns and therefore, they may not want to disclose too much information to the social planner; additionally, students may not want to possessing an excessive amount of information, especially when acquiring information is both pecuniarily and cognitively costly.

In this paper, we address the question of the exact amount of information required for implement a desired allocation. To do this, we revisit the novel concept *informational size* coined by Mount and Reiter (1974). This concept originally quantifies how much information is necessary to lead to a market equilibrium in a decentralized economic system, as different market mechanisms essentially vary in the level of information requirement to allocate resources.[2] Additionally, Even if two mechanisms output the same resource allocation, one may need more information than the other. Such difference in information requirements highlights an important but overlooked analytic aspect in the extant literature on matching mechanisms; that is, distinguishing mechanisms based on the level of information to allocate scarce resources. We therefore ask about the amount of information regarding schools' priorities and other students' preference length to *secure* a certain pairwise assignment of a student to a school, and refer to the sufficient and necessary amount of information for this purpose as the *informativeness* for a given school choice problem. Note that the original concept of informational size in Mount and Reiter (1974) serves *locally*, as it specifies the set of information message space to a particular market equilibrium outcome. In the context of school choice problems, we revise this concept to a *global* version by measuring the informational size of a matching mechanism as its informativeness in the *worst-case scenario* across all possible problems.[3]

---

[1] For example, when a dictatorship rule is adopted, the social planner only need to know the first dictator's information about her most preferred school to determine her assignment; similarly, when considering non-strategic information exchange or communication among students, it is often assumed that each student is fully aware of all the information of other students.

[2] The name, *informational size*, is also used to refer to different things in various articles. For instance, in McLean and Postlewaite (2002), informational size measures the degree to which a student can influence the posterior distribution of the state of nature when other students report truthfully.

[3] The worst-case approach has also been employed in the closely related auditability index concept proposed by Grigoryan and Möller (2024), as well the belief-free approach adopted by Liu (2020) and Chen and Hu (2023) to study

We term this worst-case measurement as *(maximal) informational size* and use it to differentiate the informativeness among four prominent matching mechanisms, the *immediate acceptance mechanism* (IA) (also known as the Boston mechanism), the *deferred acceptance mechanism* (DA), the *top trading cycles mechanism* (TTC), and the *serial dictatorship mechanism* (SD), in school choice problems. Our analysis uncovers two key insights. First, we find that SD, which is typically perceived as the *simplest* matching mechanism (Pycia and Troyan, 2023), demands a larger informational size compared to IA. This result nevertheless highlights a previously omitted advantage of IA in term of its information demand, which also partially explain the its popularity in real-world matching problems, even without desirable properties such as strategy-proofness and stability (Abdulkadiroğlu and Sönmez, 2003). Second, we observe that TTC demands less information compared to DA when the market contains a reasonable size of participants. The issue of comparison between TTC and DA has puzzled researchers both in theory (Gonczarowski and Thomas, 2023) and in experiment (Hakimov and Kübler, 2021). Our results address this puzzle from an informational perspective: in experiments with relatively fewer participants, agents tend to prefer DA over TTC as DA requires fewer information to secure one's allocation in all problems, while the opposite is true when the market size increases.

A mechanism with a small informational size (to secure an allocation) offers two distinct advantages for students: first, from an ex-ante perspective, students can more easily predict their allocations as they require less additional (and possibly costly) information to make accurate predictions; second, from an ex-post perspective, students sustain a lighter computational and cognitive load to verify the correctness of their allocations. This underscores how the informational size can serve as a metric to assess the verifiability and transparency of various mechanisms (Möller, 2022; Hakimov and Raghavan, 2022; Grigoryan and Möller, 2024), in the sense that the social planner can establish a more credible commitment to students by opting for a mechanism with a smaller informational size. In addition, a smaller informational size implies that students are required to disclose less of their personal information, which addresses the primary concern for privacy protection (Haupt and Hitzig, 2022).

Last, we develop a general allocation model with transfers in Appendix C, and apply our measurement of informational size for the single-item auction problems. In the case of static auction mechanisms, our findings reaffirm the previous result that the second-price auction is less informative than the first-price auction Grigoryan and Möller (2023). However, in the context of dynamic auctions, we show that the ascending auction is less informative than its descending-clock counterpart, which is contrary to results in (Li, 2017).

## 1.1 Illustrative Example

We present a simple example to illustrate the intuition behind our main results. Suppose that there are four students, A(lice), B(ob), C(harlie), and D(ave), and four schools, $o_1, o_2, o_3$, and $o_4$.

---

stability in two-sided matching markets. See Section 4.3 for further discussions.

Students have strict preferences over schools. Each school can only accept one student and has strict priorities over students. Without loss of generality, assume that Dave's preference is $R_D$ : $o_4, o_1, o_2, o_3$.

First, let us consider the serial dictatorship mechanism where the priority is randomly drawn. In this case, how can Dave predict her allocation? In particular, Dave may be concerned about whether he can receive his most preferred school $o_4$. In the worst case, Dave is ranked at the bottom, i.e., she will be the last dictator. Thus, to receive $o_4$, it must be the case that none of the other students receives it. Whenever is this event possible? For the first dictator, if her most preferred school is $o_1$, then she will surely not receive $o_4$. However, the analysis becomes tricky when Dave looks at the second dictator because Dave cannot know the second dictator's allocation by only knowing the second dictator's most preferred school: in the worst case, the second dictator's most preferred school is also $o_1$. Thus, to know that the second dictator will not receive $o_4$, Dave has to know the second dictator's second most preferred school is, say, $o_2$. Subsequently, Dave has to know the third dictator's three most preferred schools are $o_1, o_2$, and $o_3$. Overall, to know that Dave receives her most preferred school, six units of information about other students' preferences have to be revealed (the first dictator's most preferred school, the second dictator's two most preferred schools, and the third dictator's three most preferred schools).

Next, consider the same market with the immediate acceptance mechanism. Interestingly, the analysis becomes much easier: even in the worst case, to ensure that Dave receives his most preferred school $o_4$, we only need to know the most preferred schools of the other three students; namely, we only need three units of information about other students' preferences.

As the informational size considers the worst-case scenario, the preceding arguments are valid for any realized pairwise matching outcome. Thus, the preceding example implies that to implement a pairwise matching outcome for (Dave, $O_4$), the immediate acceptance mechanism requires less information compared to serial dictatorship (see Lemma 1).

## 1.2 Related Literature

The incentive-compatible revelation mechanisms (Hurwicz, 1973; Gibbard, 1973; Dasgupta et al., 1979; Myerson, 1979) have conveniently abstracted most real-world marketplaces as a *direct* mechanism in which all participants report their true private information in equilibrium, even if participants in real-world mechanisms may not recognize the incentive-compatible conditions and deviate from the truthfully reporting consensus (Li, 2024). In the context of school choice and college admission mechanisms, empirical evidences indicate that agents do not always behave sincerely even in some well-known strategy-proof matching mechanisms (Chen and Sönmez, 2006; Calsamiglia et al., 2010; Hassidim et al., 2016; Rees-Jones and Skowronek, 2018; Artemov et al., 2023). Also, Rees-Jones and Shorrer (2023) (Section 2.2) provide an excellent survey on the experiment literature on preference misrepresentation in strategy-proof mechanisms.

One theoretical explanation of the preceding sub-optimal behavior comes from the growing perception that participants may not have enough faith on the principal (designer)'s commitment; in other words, following the mechanism rules may not be incentive compatible for the principal. Akbarpour and Li (2020) first analyze this principal-incentive compatibility concern under mechanisms with transfers, and compare the level of credibility among different static and dynamic auction mechanisms. Möller (2022) considers the designer's commitment problem in allocation mechanisms without transfers, and shows that the designer can safely (i.e., with no objection from the agents) deviate from his commitment under the serial dictatorship mechanism. Hakimov and Raghavan (2022) further investigate that under which conditions that agents can confidently verify both the outcome and the use of the promised mechanism in school choice problems. Decerf et al. (2024) introduce the incontestability concept which requires that for a given information structure, students should not have any legitimate complaints for their assignments.

Our measurement of *informational size* in is related to this aspect in the sense that when a student wants to verify her allotment, she has to have enough information to help her to do so. In this direction, our informational size is akin to the concept of Grigoryan and Möller (2024)'s auditability index. The primary distinction lies in their approach to student types. Their index treats each student's type as an atomic unit, and for detecting any possible deviations one needs to communicate her entire preference list with other students. By contrast, we believe that a type can be more finely represented in the sense that a smaller informational size means that less information is needed to verify one's allocation outcome (or detect deviations) for a given matching mechanism. Therefore, instead of reporting her entire preference list, one can choose to communicate the information of a partial preference list to ensure an allocation to the school she desired.

Privacy protection is a significant concern in economic design (Haupt and Hitzig, 2022). Specifically, students may be concerned that their private information could be exploited in the future to their disadvantage. Our measurement can play a valuable role in addressing these concerns. It reveals both the necessary and sufficient amount of information required to implement a specific allocation. In other words, our measurement precisely identifies the level of private information that students need to disclose to the mechanism operator to achieve their preferred outcome. If they focus solely on a particular allocation, providing additional information will not lead to a better outcome. This insight can help alleviate privacy concerns and encourage students to disclose the necessary information without fear of exploitation. Accordingly, our method is also related to a small but rapidly growing literature on information acquisition in school choice (Immorlica et al., 2020; Artemov, 2021; Chen and He, 2021, 2022; Koh and Lim, 2022; Hakimov et al., 2023). Based on our measurement of informational size, a student (or a school) will not spend more than required on physical or mental effort to acquire information beyond what is necessary for guaranteeing a certain allocation outcome.

## 2 Model

### 2.1 Preliminary Definitions

We define a *school choice problem* as a 5-tuple $\Gamma = (N, O, R, \succ, q)$ with $N \equiv \{i_1, \ldots, i_n\}$ the finite set of students, $O \equiv \{o_1, \ldots, o_m\}$ the finite set of schools, $R \equiv (R_i)_{i \in N}$ the list of strict preferences over $O$, $\succ \equiv (\succ_o)_{o \in O}$ the list of strict priorities over $N$, and $q \equiv (q_o)_{o \in O}$ the set of each school's capacity. Assume that every school's priority is *substitutable* [4] Although introducing scarcity would not change our results, we assume that $2 < |N| = n \leq |\sum_{o \in O} q_o| = m$ for simplicity. Let $\mathscr{P}$ be the set of all problems.

A *matching* $x$ is a mapping from $N \cup O$ to the subsets of $N \cup O$ such that, for all $i \in N$ and $o \in O$: (i) $x(i) \in O \cup \{i\}$; (ii) $x(o) \subseteq N$ and $|x(o)| \leq q_o$; (iii) $x(i) = o$ if and only if $i \in x(o)$. That is, a matching specifies the school where each student is assigned or matched with herself, and the set of students assigned to each school. Let $X$ be the set of all matchings. For simplicity, for each $x \in X$ and $i \in N$, relabel $x_i \equiv x(i)$; for each $S \subseteq N$, denote $x_S \equiv x(i)_{i \in S}$ a *partial matching* of $x$. In the following, for each $x \in X$ and each $x_i$, we express $x_i$ as a *pairwise matching outcome* $(i, o)$ with $o = x_i$, which represents the smallest unit of partial matchings.[5]

A *mechanism* $f$ is a function that produces a matching $f(\Gamma)$ for each problem $\Gamma$; also for the pairwise matching outcome $(i, o)$, we have $f_i(\Gamma) = o$.[6] In this paper, we focus on four commonly used matching mechanisms, the immediate acceptance mechanism (IA), the deferred acceptance mechanism (DA), the top trading cycles mechanism (TTC), and the serial dictatorship mechanism (SD). We relegate the definitions of these four mechanisms to Appendix A. Also, when we refer to a mechanism $f$, we mean that $f$ is one of these four specified matching mechanisms.

### 2.2 Measurements

For each $R_i$ (resp. $\succ_o$), let $R_i^\ell$ (resp. $\succ_o^\ell$) be the corresponding order restricted to the $\ell$-th element, where $\ell$ is a natural number $\ell \leq m$. For instance, $R_i : a, b, c, d$ and $R_i^3 : a, b, c$. The *informativeness* of each $R_i^\ell$ is counted by its superscript $\ell$. For instance, $R_i^3 : a, b, c$ contains three units of information. We say $R_i^\ell$ (resp. $\succ_o^\ell$) is a *partial preference (resp. priority)* of $R_i$ (resp. $\succ_o$) as it contains partial information from $R_i$ (resp. $\succ_o$).

---

[4] A school $o \in O$ has substitutable priority if for any $N' \subset N$, $n' \in N'$, and $n'' \in N - n'$, if $n' \in Ch_o N'$, then $n' \in Ch_o N' - n''$ as well. Also, note that although the school choice problems we considered only permit that students can behave strategically. Our setting can also incorporate other two-sided markets such as college admissions, or (one-to-one) marriage markets.

[5] It is natural to look at the amount of information needed for pairwise matching outcomes, as each student is only concerned about their own matching outcome. Also, relaxing the setting to larger partial matchings involves with more than two students would not alter our main insights but considerably increase the notation burden. For instance, when we consider a partial matching that includes two pairs, e.g., $((i, o), (j, o'))$, we need compute its informational size as the sum of: (i) the informational size of one pairwise matching outcome $(i, o)$ in the original problem, and (ii) the informational size of another pairwise matching outcome $(j, o')$ in the reduced problem where $(i, o)$ is excluded.

[6] We sometimes use $f(\Gamma)$ and $x$ interchangeably to represent the matching outcome in problem $\Gamma$, if no confusion arises.

Denote $A = ((R_i^{\ell_i})_{i \in N}, (\succ_o^{\ell_o})_{o \in O})$ a *partial problem* of $\Gamma$, where $\ell_i$ (resp. $\ell_o$) represents the partial preference (resp. priority) of student $i$ (resp. school $o$). We measure the *informativeness* of each partial problem $A$ by counting the sum of its informativeness of all relevant preferences and priorities, i.e., $\sum_{i \in N} \ell_i + \sum_{o \in O} \ell_o$.[7] Let

$$I(A) \equiv \sum_{i \in N} \ell_i + \sum_{o \in O} \ell_o.$$

For each problem $\Gamma$, we say a partial problem $A$ *secures* the pairwise matching outcome $(i, o)$ under $f$, if by inputting $A$, $f$ guarantees that student $i$ will be assigned to school $o$, and vice versa. Denote $\mathscr{A}_{i,o}^f(\Gamma)$ the set of partial problems such that $f$ can secure $(i, o)$ in problem $\Gamma$. Also, fix a pair $(i, o)$ and a mechanism $f$, let $\mathscr{P}_{i,o}^f \subseteq \mathscr{P}$ be the support of problems that implements $(i, o)$; i.e., for each $\Gamma \in \mathscr{P}_{i,o}^f$, we have $f_i(\Gamma) = o$.

Given a pairwise matching outcome $(i, o)$, its *informational size* under mechanism $f$ is measured by the *maximal* informativeness among the corresponding partial problems

$$IS(f; i, o) = \max\{I(A) \mid \Gamma \in \mathscr{P}_{i,o}^f, A \in \mathscr{A}_{i,o}^f(\Gamma)\},$$

that is, from the perspective of a student $i$, $IS(f; i, o)$ measures the maximal of additional information she needs in order to secure her assignment in school $o$.

When comparing two mechanisms, say $f$ and $g$, we say $g$ is *weakly less informative* than $f$ if for each pair $(i, o)$, $IS(f; i, o) \leq IS(g; i, o)$, denoted as $IS(f) \dot{\leq} IS(g)$. If a mechanism $g$ demands more information to secure the pairwise allocation $(i, o)$ compared to another mechanism $f$; in other words, under $f$, student $i$ can more easily foresee her allocation ex-ante and verify her allocation ex-post. Also, if for each pair $(i, o)$, $IS(f; i, o) \leq IS(g; i, o)$ and for some pair $(j, o')$, $IS(f; j, o') < IS(g; j, o')$, then $g$ is *(strictly) less informative* than $f$ with $IS(f) \dot{<} IS(g)$. Also, when $IS(f) \dot{\leq} IS(g)$ and $IS(g) \dot{\leq} IS(f)$, we have $IS(f) \dot{=} IS(g)$.

## 3 Main Results

We first convey the main insights and intuition of our results through a simplified *top-ranked* school choice problem, in which for a pairwise matching outcome $(i, o)$, student $i$ ranks school $o$ as her most preferred school; i.e.,

$$\hat{R}_i : o, \ldots.$$

This step is without loss of generality from the perspective of students. For instance, if a student receives school $o$ by reporting any preferences, then certainly she will receive the same school by reporting a new preference in which school $o$ is shifted up at the top.[8] Later, we will relax this limitation and generalize the results of Lemma 1 in more general problems with heterogeneous preferences.

---

[7] In other words, our measurement is *identity-invariant* in the sense that we only count all the relevant rankings of schools' priorities which could affect student $i$'s assignment, but ignore the exact identity of each student.

[8] This is referred to as *individual monotonicity* (Takamiya, 2001). Loosely speaking, it suggests that if an agent $i$ receives $o$ at an individually monotonic mechanism $f$ by reporting $R_i$, then if she report is changed to $R_i' : o, \ldots$, ce-

Given $(i,o)$, $f$, and the corresponding support $\mathscr{P}_{i,o}^f$, let $\hat{\mathscr{P}}_{i,o}^f \subseteq \mathscr{P}_{i,o}^f$ be the support such that $i$ positions $o$ at the top (i.e., for each $\Gamma \in \hat{\mathscr{P}}_{i,o}^f$; and $\hat{R}_i : o,\ldots$). The corresponding informational size for this restricted support is counted by

$$\hat{IS}(f) = \max\{I(A) \mid \Gamma \in \hat{\mathscr{P}}_{i,o}^f, A \in \mathscr{A}_{i,o}^f(\Gamma)\}.$$

**Lemma 1.** *If $n \geq 4$, then*

(i) $\hat{IS}(IA) \lessdot \hat{IS}(SD) \lessdot \hat{IS}(TTC)) \lesseqgtr \hat{IS}(DA)$; *otherwise*

(ii) $\hat{IS}(IA) \lessdot \hat{IS}(SD) \lessdot \hat{IS}(DA) \lessdot \hat{IS}(TTC)$.

*Proof.* See Appendix B.1. □

The results in Lemma 1 can be split into two parts. First, when we look at the restricted support, SD is *less informative* than IA. This may seem counterintuitive, as SD is usually perceived as the mechanism requiring the least amount of information. For instance, Grigoryan and Möller (2024) introduce the concept of *(worst-case) auditability index* to quantify the severity of the designer's deviation incentives from executing the promised mechanism; they show that both IA and SD have an auditability index of two, while the auditability index of DA is equal to the number of students. Part (i) of Lemma 1 further distinguish the difference of informational demands in verifying the matching outcome between IA and SD. Intuitively, even if student $i$ positions her final assigned school $o$ at the top, $i$ still needs to consider the possibility that she becomes the last dictator under SD, which induces a higher information demand compared to the case under IA: Note that, for the worst case, if $i$ ranks $o$ at the top, then $i$ is matched with $o$ at step 1 under IA, and is matched with $o$ at step n under SD.

Second, the comparison between TTC and DA depends on the population size of students: when the size is small (i.e., with less than four students), TTC is *weakly less informative* than DA; however, as the population increases, DA becomes less informative. We introduce an example with three students and three schools (each with capacity one) in the proof of Lemma 1, which exhausts all the possible worst cases for a problem with less than four students, to demonstrate the the ranking reverse between TTC and DA in problems with a small population size. This result essentially comes from the absence of any possible cyclic priority structures under these two mechanisms (Ergin, 2002; Kesten, 2006), which requires at least three different students to initiate a chain of rejections and acceptances.

---

teris paribus, he will still receive $o$ at $f$. Takamiya (2001) shows that all strategy-proof mechanisms satisfy individual monotonicity, and Kojima and Ünver (2014)'s result implies that IA also satisfy individual monotonicity.

Next, we relax the restricted support of the modified top-ranked problems to the *common-priority* scenario, i.e.,

$$\succ_o = \succ_{o'}, \quad \forall o, o' \in O.$$

Note that comparing SD with other matching mechanisms is reasonable under the preceding restricted top-ranked support, as for each mechanism we only need to count the priority of one specific school; that is, we only need to take $\succ_o$ into account for IA, DA, and TTC, while focusing on the priority of the current dictator under SD. For school choice problems with common priorities, we can still compare SD with the other three mechanisms by additionally requiring that SD is executed with respect to the common priority; that is, the order of the dictators is determined by their positions in schools' common priority. Denote SD that respects the common priority $SD^\succ$.

**Theorem 1.** *In the case of common priority, $IS(IA) \dot{<} IS(SD^\succ) \dot{=} IS(TTC) \dot{=} IS(DA)$.*

*Proof.* See Appendix B.2. □

The equivalence between $SD^\succ$, TTC, and DA of Theorem 1 is not surprising. It is known that TTC and DA are identical when the common priority is imposed (Kesten, 2006), while the equivalence between $SD^\succ$ and DA is satisfied by the definition of $SD^\succ$. What is intriguing is that SD is still *strictly less informative* than IA in the case of common priority. This result once again demonstrates why IA is widely used in practical school choice problems, even though it does not satisfy certain desirable properties, such as stability and strategy-proofness.

However, when priorities are heterogeneous, it is not reasonable to compare SD with the other three mechanisms. This is because all the other three mechanisms are respect to the entire priority profile $\succ$, while SD is not. In Appendix C, we consider a general framework which includes the nature as a student who decides the realizations of the states for all other students.

**Theorem 2.** *In the case of heterogeneous priorities,*

*(i) $IS(IA) \dot{<} IS(TTC)$ and $IS(IA) \dot{<} IS(DA)$.*

*(ii) If $n \geq 4$, then $IS(TTC) \dot{<} IS(DA)$; otherwise $IS(DA) \dot{<} IS(TTC)$.*

*Proof.* See Appendix B.3. □

Theorem 2 generalized the results to an environment allowing heterogeneous priorities. As already mentioned, in this case we only compare IA with TTC and DA. Its appealing part comes from the second result, which identifies the distinct information demands between DA and TTC under different market sizes.

In experimental studies, there is no consensus on the performance between DA and TTC in terms of the ratio of truthful reporting; see a recent review by (Hakimov and Kübler, 2021).[9] Our second result in Theorem 2 partially answers this puzzle by indicating that the comparison between DA and TTC may depend on the size of population.[10] Some experimental evidence supports our findings. For instance, in Guillen and Veszteg (2021), the population of students is four and the result is that DA outperforms TTC in term of the truthtelling rate; while in Pais et al. (2011), the population of students is five and the result is that TTC outperforms than DA.

---

[9] We believe that the experimental comparison between DA and TTC can be based on the ratio of truthful reporting. Because under both mechanisms, truthful reporting never harms students, if one of them is less informative, then students under this less informative mechanism will demand more information to find that their weakly dominant strategy is truthful reporting, and therefore report truthfully.

## 4 Discussion

### 4.1 Relations with Manipulation Vulnerability

Pathak and Sönmez (2013) propose the concept of *manipulation vulnerability*, and show that: i) a variant of IA, the pre-2009 Chicago's public schools mechanism, is *more vulnerable to manipulation* than SD for any given length of students' preferences; ii) a hybrid of IA and DA, the first-preference-first mechanism, is more vulnerable to manipulation than DA for any given length of students' preferences. Their results stand in stark opposition to our informational size rankings for IA, SD, and DA, as demonstrated in Theorem 2. Nevertheless, we believe that both Pathak and Sönmez (2013)'s manipulation vulnerability concept and our informational size concept adhere to a unified economic rationale: when a student demands less information to foresee a secured outcome of her manipulation, she is evidently more inclined to undertake strategic behavior under this mechanism.

In addition, Pathak and Sönmez (2013) and Chen and Kesten (2017) find that both DA and a hybrid of IA and DA, the Chinese parallel mechanism, become more vulnerable to manipulation, when we shorten the length of students' preference. The following proposition, nevertheless, reveals that a shortened preference length also induces less information burden on students, which also contrasts to the ranking to students' manipulation incentives demonstrated in the previous two papers. This is also because a student clearly requires less information to achieve her manipulations when she includes fewer schools in her preference list, as the economic rationale mentioned above.

**Proposition 1.** *Suppose there are at least $\kappa$ schools, and $\kappa > e > e' > 0$. Then $IS(f^{e'}) \dot{<} IS(f^e)$, $f \in \{IA, SD, TTC, DA\}$.*

Proposition 1 reveals an important but overlooked advantage of shortening students' preference length from the perspective of information demands, especially when acquiring extra information may incur either pecuniary or cognitive costs. We omit the proof of Proposition 1, as it directly follows the proof of Theorem 2 with $\kappa < n \leq m$.

---

[10]Although our result presents a threshold value of four students, we do not intend to interpret it as the exact cutoff for the discrepant information demands between DA and TTC in the real world. Instead, it serves more as an illustrative example to reveal the importance of population size when a student needs to forecast and compute the behavior of other students.

## 4.2 Most-informative Mechanisms

We further discuss some possible characterizations on determining the *most-informative* mechanisms, and consider within the classes of strategy-proof mechanisms and efficient mechanisms. We say a mechanism $f$ is *most-informative* among a set of mechanisms $\mathcal{F} (\ni f)$, if for each $g \in \mathcal{F} \setminus \{f\}$, $g$ is *weakly less informative than* $f$.

**Strategy-proof mechanisms.** When we focus on this class, it is easy to see that the constant mechanisms are *most-informative* as their output does not depend on the preference or priority profile.[11]

**Efficient mechanisms.** It is known that IA, TTC, and SD are *Pareto efficient*.[12] Additionally, our results show that given that TTC is *less informative than* IA and SD, it leads to the question of whether IA or SD is *most-informative* among all *Pareto efficient* mechanisms. Our answer is no. The reason is straightforward: Pareto efficiency is solely determined by students' preferences, whereas IA considers both the preference profile and the priority profile, and SD takes the exogenous moving sequence into account. Thus, it is conceivable that we can identify an *efficient* mechanism that only considers the preference profile. We provide such an example below.

**Example 1.** The idea here is that we can use one student's preferences to generate a moving sequence in which this particular student consistently ranks at the bottom.

Let $O' = \{o_1, \ldots, o_{n-1}\}$. For each $R_n$, let $R_{n|O'}$ be the corresponding order of $R_n$ restricted to $O'$, and $o(i, R_{n|O'}) \in O'$ be the $i$-th preferred school among $O'$ at $R_{n|O'}$. Let $\rho : O' \to \{1, \ldots, n-1\}$ be a bijection such that for each $i \in \{1, \ldots, n-1\}$, $\rho(o_i) = i$. Next, let $\rhd : N \to N$ be a common priority that is induced by $R_n$ such that

- For each $i \in \{1, \ldots, n-1\}$, $\rhd(i) = \rho(o(i, R_n))$; and

- $\rhd(n) = n$.

For instance, if $R_n : o_2, o_1, o_3, o_4, \ldots, o_{n-1}, o_n, \ldots$ then $\rhd : 2, 1, 3, 4, \ldots, n-1, n$.

For each problem, we execute IA with respect to the induced common priority $\rhd$, and denote this mechanism by $IA^{\rhd}$. It is easy to see that $IA^{\rhd}$ is *Pareto efficient* as it is a modification of IA. ◇

Thus, it is natural to conjecture that $IA^{\rhd}$ is *most-informative* among all *Pareto efficient* mechanisms. That conjecture is almost correct; however, we need to strengthen *Pareto efficiency* to *favoring-higher-ranks*. *Favoring-higher-ranks* is introduced by Kojima and Ünver (2014), it says that for each school $o$, if there is a student $i$ who prefers $o$ to her allotment, then $o$ is not assigned to a student who ranks it lower.[13]

---

[11]A mechanism $f$ is a *constant mechanism* if for any two problems $\Gamma$ and $\Gamma'$, $f(\Gamma) = f(\Gamma')$.

[12]A matching $x$ is *Pareto efficient* at $R$ if there is no another matching $y(\neq x)$ such that for each $i \in N$, $y_i R_i x_i$. A mechanism is *Pareto efficient* if it always outputs *Pareto efficient* matchings.

[13]Formally, for each $i \in N$, each $o \in O$, and each $R_i$, let $r(o, R_i)$ be the rank of $o$ at $R_i$, i.e., $r(o, R_i) = |\{o' \in O \mid o' R_i o\}|$. A mechanism $f$ satisfies *favoring-higher-ranks* if each $\Gamma \in \mathcal{P}$, each $i \in N$, and each $o \in O$, if $o P_i f_i(\Gamma)$, then there is no student $j \in N$ such that $f_j \Gamma = o$ and $r(o, R_j) > r(o, R_i)$. Proposition 1 of Kojima and Ünver (2014) shows that *favoring-higher-ranks* implies *Pareto efficiency* whenever all matchings are feasible.

**Proposition 2.** $IA^{\triangleright}$ *is the* most-informative *among all mechanisms that satisfy* favoring-higher-ranks.

*Proof.* See Appendix B.4. □

## 4.3 Robustness of the Worst-case Approach

As noted in Footnote 3, although analyzing worst-case scenarios has been widely adopted as a belief-free approach in the literature on matching with incomplete information, a cautious reader may question whether our metric of informational size is too sensitive. We believe that our metric of informational size can be robustly extended beyond the worst-case scenario. For instance, instead of focusing on the worst case, one might consider the second-worst case as a *cutoff* with a different size of relevant information to determine which mechanisms fall above or below this threshold. Although this alternative metric would quantitatively change the exact informational size for each mechanism we considered, it will not invalidate the ranking and qualitative insights we observed through the measurement under the worst-case scenario.

However, we cannot replace our pairwise matching analysis by a *problem-wise* basis, in the sense that comparing the information demands of a mechanism $f$ with another mechanism $g$ for each $\Gamma \in \mathcal{P}$. This is because two mechanisms may have different supports when predicting a particular allocation. For instance, consider a pairwise matching $(i, o)$. It might be the case that $i$ is matched with $o$ under $DA(\Gamma)$ but not under $TTC(\Gamma)$. This discrepancy arises because the support of $(i, o)$ under the two mechanisms differs. This is where the advantage of worst-case analysis comes into play. By focusing on worst-case scenarios, we can consistently compare two mechanisms based on any (partial) allocations. This approach ensures that we have a common ground for comparison, regardless of the specific disparities in the supports of different mechanisms. It also provides us a more clear-cut and consistent conceptual structure for assessing the information demands requirements of various mechanisms.

# 5 Conclusion

In this paper, we propose a novel measurement, *informational size*, and apply it to analyze the information demands discrepancies for different mechanisms in school choice problems. Apart from its theoretical implications, the clear-cut rankings of information demands among the four prominent matching mechanisms (i.e., the immediate acceptance mechanism, the deferred acceptance mechanism, the top trading cycles mechanism, and the serial dictatorship mechanism) are especially pertinent when acquiring extra information is costly for either financial reasons or cognitive constraints (or both). Our informational size metric also sheds light on: i) the discrepancies between theoretical predication and experimental evidence regarding the truth-telling rate for the deferred acceptance mechanism versus the top trading cycles mechanism; and ii) uncovering the source of auditability and manipulation vulnerability differences among some prominent matching mechanisms.

We conclude with some avenues for future research. First, as illustrated by the general framework in Appendix C, we can extend the application of our measurement to other problems, including voting rules, to gain a broader understanding of its applicability. Second, given the simple construction of our measurement, we can also conduct experiments to validate and further support our findings. Last, we have not discussed the most-informative mechanism among all *stable* mechanisms.[14] This is because the analysis of stability could be quite intricate, as some matchings are non-rationalizable in the sense of Echenique (2008); currently, we do not have an unambiguous answer to this question and defer it to future research.

# Appendices

## Appendix A  Matching Mechanisms

### A.1  Immediate Acceptance Algorithm

**Input.** A problem $\Gamma \in \mathscr{P}$.

**Step 1.** Let $N(1) := N$ and $O(1) := O$. Each student $i \in N(1)$ proposes to the most preferred school in $O(1)$ (according to $R_i$). We say that student $i$ is a current proposer of $o$. For each school $o \in O(1)$, let $N_o(1)$ be the set of $o$'s current proposers. We match each school $o$ to the student in $N_o(1)$ who has the highest priority in $o$ (according to $\succ_o$) and remove all matched students and schools. We define $N(2)$ to be the set of remaining students and $O(2)$ to be the set of remaining schools, and if $N(2)$ is not empty, we continue with Step 2. Otherwise, the algorithm terminates.

In general at Step $t$ we have the following:

**Step $t \geq 2$.** Each student $\in N(t)$ proposes to the most preferred school in $O(t)$ (according to $R_i$). We say that student $i$ is a current proposer of $o$. For each school $o \in O(t)$, let $N_o(t)$ be the set of $o$'s current proposers. We match each school $o$ to the student in $N_o(t)$ who has highest priority at $o$ (according to $\succ_o$) and remove all matched students and schools. We define $N(t+1)$ as the set of remaining students and $O(t+1)$ as the set of remaining schools, and if $N(t+1)$ is not empty, we continue with Step $t+1$. Otherwise, the algorithm terminates.

**Output.** The immediate acceptance algorithm terminates when each student is matched.

### A.2  (Student-proposing) Deferred Acceptance Algorithm

**Input.** A problem $\Gamma \in \mathscr{P}$.

---

[14]A matching $x$ is *stable* at $\Gamma$ if there is no pair $(j, o)$ with $x_j \neq o$ such that $o P_j x_j$ and $j \succ_o i$ where $x_i = o$. A mechanism is *stable* if it always outputs *stable* matchings.

**Step 1.** For each $i \in N$, let $O_i(1) := O$ be student $i$'s available set. Each student $i$ proposes to the most preferred school in $O_i(1)$ (according to $R_i$). We say that student $i$ is a current proposer of $o$. For each school $o \in O$, let $N_o(1)$ be the set of $o$'s current proposers. We tentatively match each school $o$ to the student in $N_o(1)$ who has the highest priority at $o$ (according to $\succ_o$) and rejects all other proposers. For each student $i \in N$, if she is rejected by her proposed school $o$, we define $O_i(2) = O_i(1) \setminus \{o\}$ be her available set after the rejection from $o$; otherwise $O_i(2) = O_i(1)$. If some students are rejected at this step, we continue with Step 2. Otherwise, the algorithm terminates.

In general at Step $t$ we have the following:

**Step $t \geq 2$.** Each student $i$ proposes to the most preferred school in $O_i(t)$ (according to $R_i$). For each school $o \in O$, let $N_o(t)$ be the set of $o$'s current proposers. Note that $o$'s tentative owner is also in $N_o(t)$. We tentatively match each school $o$ to the student in $N_o(t)$ who has the highest priority in $o$ (according to $\succ_o$) and rejects all other proposers. For each student $i \in N$, if she is rejected by her proposed school $o$, we define $O_i(t) = O_i(t-1) \setminus \{o\}$ be her available set after the rejection from $o$. If some students are rejected at this step, we continue with Step $t+1$. Otherwise, the algorithm terminates.

**Output.** The deferred acceptance algorithm terminates when each student is tentatively matched.

## A.3 Top Trading Cycles Algorithm

**Input**. A problem $\Gamma \in \mathscr{P}$.

**Step 1.** Let $N(1) := N$ and $O(1) := O$. We construct a (directed) graph with the set of nodes $N(1) \cup O(1)$. For each student $i \in N(1)$ we add a directed edge to her most preferred school in $O(1)$ (according to $R_i$). For each directed edge $(i, o)$, we say that student $i$ points to school $o$. For each school $o \in O(1)$ we add a directed edge to the student who has the highest priority at $o$ in $N(1)$ (according to $\succ_o$). A trading cycle is a directed cycle in the graph. Given the finite number of nodes, at least one trading cycle exists for this graph. We match to each student in a trading cycle the school she points to and remove all trading cycle students and schools. We define $N(2)$ to be the set of remaining students and $O(2)$ to be the set of remaining schools, and if $N(2)$ is not empty, we continue with Step 2. Otherwise, the algorithm terminates.

In general at Step $t$ we have the following:

**Step $t \geq 2$.** We construct a (directed) graph with the set of nodes $N(t) \cup O(t)$ where $N(t) \subsetneq N$ is the set of students that remain after Step $t-1$ and $O(t) \subsetneq O$ is the set of schools that remain after Step $t-1$. For each student $\in N(t)$ we add a directed edge to her most preferred school in $O(t)$ (according to $R_i$). For each school $o \in O(t)$ we add a directed edge to the student who has the highest priority at $o$ in $N(t)$ (according to $\succ_o$). At least one trading cycle exists for this graph, and we match to each student in a trading cycle the school she points to and remove all trading cycle students and schools. We define $N(t+1)$ to be the set of remaining students and $O(t+1)$ to be the set of remaining schools, and if $N(t+1)$ is not empty, we continue with Step $t+1$. Otherwise, the algorithm terminates.

**Output.** The top trading cycles algorithm terminates when each student is matched.

## A.4 Serial Dictatorship Algorithm

**Input.** A problem $\Gamma \in \mathscr{P}$ and a moving sequence $\pi$, which can be stated as a linear order over $N$ or a bijection from $N$ to $\{1,\ldots,n\}$.

**Step** 1. Let $SD_{\pi(1)}$ be student $\pi(1)$'s most preferred school in $O$ (according to $R_{\pi(1)}$). Match $SD_{\pi(1)}$ to student $\pi(1)$.

**Step** $t \geq 2$. Let $SD_{\pi(t)}$ be student $\pi(t)$'s most preferred school in $O \setminus \{SD^\pi_{\pi(1)}, \ldots, SD_{\pi(t-1)}\}$ (according to $R_{\pi(t)}$). Match $SD_{\pi(t)}$ to student $\pi(t)$.

**Output.** After $n$ steps, allocation $\{SD_{\pi(1)}, \ldots, SD^\pi_{\pi(n)}\}$ is determined.

# Appendix B   Omitted Proofs

We adopt the following three notation simplifications throughout the proofs, if no confusion arises.

First, since the first step in each proof is always to fix a pair $(i,o)$. Also, for each mechanism $f$, we only consider the corresponding support $\mathscr{P}^f_{i,o}$. We will omit this step henceforth.

Second, for each pair $(i,o)$ and each mechanism $f$, when we calculate $IS(f;i,o)$, we do not count $R_i$. The interpretation is that when we explain $IS(f;i,o)$ as the amount of additional information that student $i$ needs to know to confirm her allocation at $f$ is $o$, of course student $i$ knows her own private information $R_i$. Alternatively, we can count $R_i$ by adding a constant number for each mechanism. For instance, if $o$ is the $k$-th best school at $R_i$, then for each $f$, we always count $R^k_i$ when we compute $IS(f;i,o)$. Since our goal is to compare the difference between the informational size of different mechanisms, adding a constant counting factor does not affect our results. We thus omit the amount of informativeness from $R_i$ when computing the information sizes.

Third, although we consider the (many-to-one) school choice problem, in the proof we essentially adopt a cloning procedure as illustrated by Roth and Sotomayor (1990). That is, each school with capacity $q_o$ is cloned $q_o$ times, which generates $q_o$ clone schools. Each clone school has one seat, and shares the same priorities as the original school; each agent prefers the first clone than the second, and so on. Formally, for each problem $(N,O,R,\succ,q)$, we induce an unique clone problem $(N,O',R',\succ',q')$, where $O' = \{o_1^1, \ldots, o_1^{|q_{o_1}|}, \ldots, o_m^1, \ldots, o_m^{|q_{o_m}|}\}$, $q' = (1,\ldots,1)$, $(R',\succ')$ is such that for each $i \in N$,

- for $j,k = 1,\ldots,m$, and $l(\leq q_j), \ell(\leq q_k)$, $o_j^l R'_i o_k^\ell$ if and only if $o_j R_i o_k$;

- for $j = 1,\ldots,m$, and $l < \ell \leq q_j$, $o_j^l P'_i o_j^\ell$; and

- for $j = 1,\ldots,m$, and $l < \ell \leq q_j$, $\succ'_{o_j^l} = \succ'_{o_j^\ell} = \succ_{o_j}$.

Therefore, instead of computing the informational size of the original problem, we can focus on the corresponding clone problem.

Two things to be noted for the clone problem. First, for each pairwise matching outcome $(i,o)$ at the original problem, when we look at the worst case, it should be that $i$ is matched with the last seat, hence, at the clone problem, it will be $(i, o^{|q_o|})$. However, when we compute the informational size for it, it is conditional on the fact that $o^1, \ldots, o^{|q_o|-1}$ are matched with other agents. Therefore, our analysis below also works. For instance, if $o$ is $i$'s first best school, then at the clone problem, we ask whenever $i$ is matched with his $|q_o|$ best school, $o^{|q_o|}$. Second, when we compute the informational size for the clone problem, actually we redundantly compute something. For instance, after compute the informativeness of $\succ'_{o^1}$, we automatically know $\succ'_{o^2}$ as well. Our analysis below compute both of them. Therefore, there is some redundant informativeness. However, this will not change our result as our goal is to compare different mechanisms. That is, if for a mechanism $f$ we compute some redundant informativeness, such as the informativeness of $\succ'_{o^2}$, for another mechanism $g$ we also redundantly compute this amount of informativeness. Therefore, when we compare the informational size between $f$ and $g$, those amount of redundant informativeness will be double canceled.

## B.1 Proof of Lemma 1

In terms of $\hat{IS}(f; i, o)$, it is easy to see that for the worst case, we must have either (i) $\succ_o: \ldots, i$ for IA and TTC, or (ii) $\pi: \ldots, i$ for SD, where $\pi$ represents a given moving sequence of dictators as illustrated in Appendix A.4; i.e., student $i$ has the lowest priority at $o$ or moves lastly under $\pi$. Thus, for $f \in \{IA, TTC, SD\}$, $IS(f; i, o) \geq n-1$ as student $i$ at least to know $\succ_o^{n-1}$ or $\pi^{n-1}$ (once we know $n-1$ elements we automatically know the full list). However, in the case of DA, knowing $\succ_o$ is unnecessary: agent $i$ always has a chance to be tentatively matched with school $o$, regardless of $o$'s priority, see Example 2 below for details.

Recall that $\hat{R}_i : o, \ldots$, for each mechanism $f$, we focus on the support $\hat{\mathcal{P}}_{i,o}^f$ in the proof of Lemma 1.

**IA**

By definition, we can see that student $i$ will be matched to $o$, if there is no another student positions $o$ at the top. Therefore, in the worst case (i.e., student $i$ has the lowest priority at $o$ or moves lastly under $\pi$), student $i$ needs to know $(\hat{R}_j^1)_{j \in -i}$ such that $\hat{R}_j^1 : o'_j, \ldots$, for each $j \in -i$, $o'_j \neq o$.[15] Thus, the corresponding partial problem is $((\hat{R}_j^1)_{j \in -i}, \succ_o^{n-1})$. Hence, we have

$$\hat{IS}(IA; i, o) = (n-1) + (n-1) = 2(n-1).$$

**SD**

---

[15] In contrast, if there is a student $j$ with $\hat{R}_j^1 : o, \ldots,$, then $f_i(\Gamma) \neq o$ as $j \succ_o i$ and $j$ also proposes to $o$ at Step 1.

Consider a problem $\Gamma \in \mathscr{P}_{i,o}^f$. For SD with a corresponding moving sequence $\pi$, student $i$ will have the last move in the worst case. Thus, student $i$ at least knows $\pi^{n-1}$. For the first dictator $\pi(1)$, it is sufficient to know that $\pi(1)$'s first choice, i.e., $\hat{R}_{\pi(1)}^1 : o'(\neq o),\ldots$; otherwise, $SD_{\pi(1)}(\Gamma) = o$. Subsequently, conditional on $\hat{R}_{\pi(1)}^1$, for the second dictator $\pi(2)$, in the worst case, it is sufficient to know that her first and second choices are not $o$, i.e., $\hat{R}_{\pi(2)}^2 : o'(\neq o), o''(\neq o)\ldots$; otherwise, $SD_{\pi(2)}(\Gamma) = o$. Thus, it is sufficient to know $(\hat{R}_{\pi(1)}^1, \hat{R}_{\pi(2)}^2, \ldots, \hat{R}_{\pi(n-1)}^{n-1})$ and $\pi^{n-1}$ to secure $(i, o)$ in the worst case. We have

$$\hat{IS}(SD; i, o) = [1 + 2 + 3 + \ldots (n-1)] + (n-1) = \sum_{\ell=1}^{\ell=n-2} \ell + (2n-2) = \frac{(n-1)(n-2)}{2} + 2(n-1).$$

**TTC**

Without loss of generality, let $\rho_1$ be the student who has the highest priority at $o$, $\rho_2$ the student who has the second highest priority at $o$, and so on. By the definition of TTC, in the worst case, we have the following steps: at each step, there is only one pairwise trading cycle. In particular, at each step $\ell$, we assume that all remaining students (except for $i$) point to $o_\ell = TTC_{\rho_\ell}(\Gamma)$ and all remaining schools point to $\rho_\ell$. Thus, $\rho_\ell$ is matched with $o_\ell$. As a result, at each step $\ell$, it is necessary to know $\hat{R}_{\rho_\ell}^\ell$ and $\succ_{o_\ell}^\ell$. So, the corresponding partial problem is $((\hat{R}_{\rho_\ell}^\ell, \succ_{o_\ell}^\ell)_{\ell=1}^{\ell=n-1}, \succ_o^{n-1})$. Thus,

$$\hat{IS}(TTC; i, o) = n-1+(1+1)+(2+2)+\ldots+(n-1+n-1) = n-1+2\sum_{\ell=1}^{n-1} \ell = n-1+n(n-1) = (n+1)(n-1).$$

**DA**

The preceding arguments have revealed that $\hat{IS}(IA) \dot{<} \hat{IS}(SD) \dot{<} \hat{IS}(TTC)$, for any $n \geq 2$. Thus, it is suffices to show that there is a problem $\Gamma$ and a corresponding partial problem $A \in \mathscr{A}_{i,o}^f(\Gamma)$ with $\hat{IS}(DA; i, o) \geq (n+1)(n-1)$.

To show that $\hat{IS}(TTC) \dot{\leq} \hat{IS}(DA)$ when $n \geq 4$, we need to construct the worst case under DA, which generates as many rejections as possible. Without loss of generality, relabel $i = n$ and $o = o_n$, i.e., $n$ has the lowest priority at $o_n$. Consider the (worst) case as follows:

> *Step* 1: student 1 proposes to $o_2$, and each student $\ell \neq 1$ proposes to $o_\ell$. Moreover, $1 \succ_{o_2} 2$; thus, student 2 is rejected.
>
> *Step* 2: student 2 proposes to $o_3$, and $2 \succ_{o_3} 3$; thus, student 3 is rejected.
>
> $\vdots$
>
> *Step* $n-1$: student $n-1$ proposes to $o_2$, and $n-1 \succ_{o_2} 1$; thus, student 1 is rejected.
>
> *Step* $n$: student 1 proposes to $o_3$, and $1 \succ_{o_3} 2$; thus, student 2 is rejected.
>
> $\vdots$
>
> *Step* $n(n-2)$: student $n-1$ proposes to $o_{n-1}$, and $n-1 \succ_{o_{n-1}} 1$; thus, student 1 is rejected.

*Step $n(n-2)+1$:* student 1 proposes to $o_1$, since no one proposes it, student 1 is matched. Finally, the outputted matching is for each $\ell \in N$, $DA_\ell(\Gamma) = o_\ell$.

We find that at each step, there is one pairwise comparison, which means that to output $(n, o_n)$, we must know at least two units of information: the active student's proposal and her priority at the proposed school. The corresponding partial problem is $(\hat{R}_\ell^{n-2}, \succ_{o_\ell}^{n-1})_{\ell=1}^{n-1}$, and its informational size is

$$\hat{IS}(DA; i, o) = (n-2)(n-1) + (n-1)(n-1) = (2n-3)(n-1).$$

Thus, $\hat{IS}(DA; i, o) \geq \hat{IS}(TTC; i, o)$ if and only if $2n - 3 \geq n + 1$, which gives us $n \geq 4$. Since the preceding arguments are valid for any pair $(i, o) \in N \times O$, we have $\hat{IS}(DA) \dot{\geq} \hat{IS}(TTC)$ when $n \geq 4$.

For the case when $n < 4$, we introduce the following example, which exhausts all the possible worst cases with three students, to show that $\hat{IS}(DA) \dot{<} \hat{IS}(TTC)$.

**Example 2.** Let $N = \{1, 2, 3\}$, $O = \{a, b, c\}$, and $q_{o \in O} = 1$. Without loss of generality, we construct the worst case for student 1 under TTC as:

| $\succ_a$ | $\succ_{o,\, o=b,c}$ | $R_1$ | $R_2$ | $R_3$ |
|---|---|---|---|---|
| 2 | 3 | [a] | b | [b] |
| 3 | 2 | b | [c] | c |
| 1 | 1 | c | a | a |

Each squared box represents the assignment of each student. Clearly, for this problem $\Gamma$, we have the corresponding partial problem $A_{1,a} = (R_2^2, R_3^1, \succ_a^2, \succ_b^1, \succ_c^2)$, with $A \in \mathcal{A}_{1,a}^{TTC}(\Gamma)$ and $I(A) = 8$. Thus, when $n = 3$, we have $\hat{IS}(TTC; 1, a) = 8$. Since we can construct a parallel partial problem for any pair $(i, o) \in N \times O$ as above, we have $\hat{IS}(TTC) = 8$.

Next, we construct the worst case under DA, with the corresponding partial problem is $(R_2^2, R_3^1, \succ_a^2, \tilde{\succ}_b^2, \succ_c^0)$. This gives us $\hat{IS}(DA; 1, a) = 7$. Since we select student 1 and school $a$ in an arbitrary way, we can construct a similar partial problem for any pair $(i, o) \in N \times O$ as above. Thus, we have $\hat{IS}(DA) = 7$.

| $\succ_a$ | $\tilde{\succ}_b$ | $\succ_c$ | $R_1$ | $R_2$ | $R_3$ |
|---|---|---|---|---|---|
| 2 | 1 | 3 | [a] | b | [b] |
| 3 | 3 | 2 | b | [c] | c |
| 1 | 2 | 1 | c | a | a |

◇

## B.2 Proof of Theorem 1

In the case of common priority, it is known that TTC, DA, and SD coincide with each other. Thus, $IS(TTC) \dot{=} IS(DA) \dot{=} IS(SD^\succ)$. Moreover, from the proof of Lemma 1, it is easy to see that for each pair $(i, o)$, $IS(SD^\succ; i, o) = \frac{(n-1)(n-2)}{2} + 2(n-1)$, regardless of the $o$'s position at $R_i$.

So, we only need to show that $IS(IA) \dot{<} IS(SD^\succ)$. Consider $o$'s position at $R_i$ with the following cases.

If $o$ is the first-best school at $R_i$. By the analysis in the proof of Lemma 1, we have

$$IS(IA; i, o) = (n-1) + (n-1) = 2(n-1).$$

In this case, $IS(IA; i, o) < IS(SD^\succ; i, o)$.

If $o$ is the second-best school at $R_i$. It is easy to see that to construct the worst case under IA, we need as many rejections as possible. Without loss of generality, assume all students propose the same school; thus, after Step 1, only one student is matched. Thus, we need to know all students but $i$'s first choice. Moreover, we additionally need to know the common priority to identify that student $i$ has the lowest priority. Without loss of generality, label the matched pair be $(1, o_1)$. To output $i$ is matched with $o$ at Step 2, in the worst case, it is sufficient to know that: (i) student $i$ has the lowest priority at $o$, and (ii) all remaining students but $i$'s second choice is not $o$. The corresponding partial problem is $A_2^{IA} \equiv ((R_1^1, (R_j^2)_{j \in N \setminus \{1, i\}}), \succ^{n-1})$. Similarly, we use $A_\ell^f$ the partial problem under mechanism $f$ when $o_n$ is ranked as the $\ell$'s best school in $R_i$. Thus, if $i$ ranks $o$ at the second, then the corresponding informational size under IA is the informativeness of $A_2^{IA}$, i.e., $I(A_2^{IA})$. That is,

$$IS(IA; i, o) = I(A_2^{IA}) = 1 + 2(n-2) + (n-1) = 2(n-1) + (n-2).$$

In this case, $IS(IA; i, o) < IS(SD^\succ; i, o)$.

The above analysis can be easily extended to the case where $o$ is the third best school at $R_i$ and so on. In the end, when $o$ is the last best school at $R_i$, the informational size is

$$IS(IA; i, o) = I(A_3^{IA}) = 2(n-1) + (n-2) + (n-3) + \ldots + (n-n) = n(\frac{n-1}{2}) + (n-1).$$

We see that $n(\frac{n-1}{2}) + (n-1) = \frac{(n-1)(n-2)}{2} + 2(n-1) = IS(SD^\succ; i, o)$. Since the preceding arguments is applicable for any pair $(i, o) \in N \times O$, we have $IS(IA) \dot{<} IS(SD^\succ)$.

## B.3 Proof of Theorem 2

Recall that in the worst case (when $i$ has lowest priority at $o$), it is necessary to know $\succ_o^{n-1}$. Also, matching $i$ with $o$ is conditional on the event that student $i$ is not matched with another school $o'$ that she prefers to $o$. Thus, it is also necessary to know $i$'s priority at $o'$ to verify that $o'$ is matched with another student: in the worst case (when $i$ has lowest priority at $o'$), we need the information about $\succ_{o'}^{n-1}$.

Without loss of generality, relabel $i = n$ and $o = o_n$, i.e., $n$ has the lowest priority at $o_n$. Also, given $R_n$, for each $k \in \{1, \ldots, n\}$, let $o^k$ be the $k$-th best school at $R_n$.

**IA**

If $o_n$ is the first-best school at $R_n$. By the analysis of IA in Lemma 1, in the worst case, the corresponding partial problem is $A_1^{IA} \equiv ((R_j^1)_{j \in -n}, \succ_o^{n-1})$, and we have

$$IS(IA; n, o_n) = I(A_1^{IA}) = (n-1) + (n-1) = 2(n-1).$$

If $o_n$ is the second-best school at $R_n$. By the analysis of IA in Theorems 1 and 1, in the worst case, the corresponding partial problem is $A_2^{IA} \equiv (R_{\rho^1}^1, (R_j^2)_{j \in N \setminus \{\rho^1, n\}}, \succ_{o^1}^{n-1}, \succ_{o^2}^{n-1})$, where $\rho^1$ is the student who is matched with $o^1$. Thus,

$$IS(IA; n, o_n) = I(A_2^{IA}) = 1 + 2(n-2) + 2(n-1).$$

The preceding arguments can be extended to the case in which $o_n$ is the third best school at $R_n$, and so on. Recursively, we have

$$I(A_k^{IA}) = I(A_{k-1}^{IA}) + [(n-1) + (n-1-k)], \quad \text{for } k = 2, \ldots, n-1, \text{ and}$$

$$I(A_n^{IA}) = I(A_{n-1}^{IA}).$$

Note that the last equation is from the fact that once we have $A_{n-1}^{IA}$, we know that under IA all other agents are matched, and thus at last step, only agent $n$ and $o_n$ are remaining, and $n$ can be matched with $o_n$ without any additional information.

**TTC**

If $o_n$ is the first-best school at $R_n$. By the analysis of TTC in Lemma 1, in the worst case, the corresponding partial problem is $A_1^{TTC} \equiv (\succ_{o_n}^{n-1}, (R_{\rho_\ell}^\ell, \succ_{o_\ell}^\ell)_{\ell=1}^{\ell=n-1} = (\succ_{o_n}^{n-1}, R_{\rho_1}^1, \succ_{o_1}^1, R_{\rho_2}^2, \succ_{o_2}^2, \ldots, R_{\rho_{n-1}}^{n-1}, \succ_{o_{n-1}}^{n-1})$.

Thus, we conclude that

$$IS(TTC; n, o_n) = I(A_1^{TTC}) = (n+1)(n-1).$$

If $o_n$ is the second-best school at $R_n$. By the analysis of TTC in Lemma 1, we need to know the informativeness of $A_1^{TTC}$, and $\succ_{o^1}^{n-1}$ in the worst-case scenairo. Recall that $o^1 \in \{o_\ell\}_{\ell=1}^{\ell=n-1}$, so in the worst case, when $o^1 = o_1$,[16] the corresponding partial problem is $A_2^{TTC} \equiv (\succ_{o^1}^{n-1}, R_{\rho_1}^1, \succ_{o_n}^{n-1}, (R_{\rho_\ell}^\ell, \succ_{o_\ell}^\ell)_{\ell=2}^{\ell=n-1})$, and

$$IS(TTC; n, o_n) = I(A_2^{TTC}) = (n+1)(n-1) + [(n-1) - 1] = (n+1)(n-1) + (n-2).$$

Similarly, when $o_n$ is the third-best school at $R_n$, in the worst case, when $o^1 = o_1$ and $o^2 = o_2$, the corresponding partial problem is $A_3^{TTC} \equiv (\succ_{o_1}^{n-1}, R_{\rho_1}^1, \succ_{o_2}^{n-1}, R_{\rho_2}^2, \succ_{o_n}^{n-1}, (R_\ell^\ell, \succ_{o_\ell}^\ell)_{\ell=3}^{\ell=n-1})$. So we have

$$IS(TTC; n, o_n) = I(A_3^{TTC}) = (n+1)(n-1) + (n-2) + [(n-1) - 2].$$

---

[16]This is without loss of generality, the point here is that student $n$'s first best school is matched at Step 1, and we can rename the school that is matched at Step 1 as $o_1$.

The above analysis can be easily extended to the case where $o$ is the fourth best school at $R_i$ and so on. Recursively, we have

$$I(A_k^{TTC}) = I(A_{k-1}^{TTC}) + [(n-1) - (k-1)] \text{ for } k = 2, \ldots, n-1, \text{ and}$$

$$I(A_n^{TTC}) = I(A_{n-1}^{TTC}).$$

Note that the last equation is from the fact that once we have $A_{n-1}^{TTC}$, we know that under TTC all other agents are matched, and thus at last step, only agent $n$ and $o_n$ are remaining, and $n$ can be matched with $o_n$ without any additional information.

**IA v.s. TTC**

It is easy to see that for $k = 2, \ldots, n-2$, $I(A_k^{IA}) < I(A_k^{TTC})$ and $I(A_k^{IA}) = I(A_k^{TTC})$ for $k = n-1, n$. Thus, $IS(IA) \dot{<} IS(TTC)$.

**DA v.s. TTC**

The analysis of DA in Lemma 1 implies that for each $(i, o)$, $IS(DA; i, o) \geq (2n-3)(n-1) = \hat{IS}(DA; i, o)$. Recall that $\hat{IS}(DA; i, o)$ is obtained by the corresponding partial problem $(R_\ell^{n-2}, \succ_{o_\ell}^{n-1})_{\ell=1}^{n-1}$. We restate it as $\hat{A}_n^{DA} \equiv (R_1^{n-2}, \ldots, R_{n-1}^{n-2}, \succ_{o_1}^{n-1}, \ldots \succ_{o_{n-1}}^{n-1})$. Note that we do not need to know anything about $\succ_{o_n}$.

Also, recall that
$$A_n^{TTC} = (R_{\rho_1}^1, R_{\rho_2}^2, \ldots, R_{\rho_{n-2}}^{n-2}, R_{\rho_{n-1}}^{n-1}, \succ_{o_1}^{n-1}, \ldots \succ_{o_{n-1}}^{n-1}).$$

We see that $I(\hat{A}_n^{DA}) - I(A_n^{TTC}) = [(n-2)-1] + [(n-2)-2] + \ldots + [(n-2)-(n-2)] + [(n-2)-(n-1)] = (n-2)(n-1) - \frac{(1+n-1)(n-1)}{2} = (0.5n-2)(n-1)$. Note that each term $[(n-2) - \ell]$ means $R_{\rho_\ell}^{n-2} - R_{\rho_\ell}^\ell$.

We see that when $n \geq 4$, $I(\hat{A}_n^{DA}) - I(A_n^{TTC}) \geq 0$. Thus, we confirm that if $n \geq 4$, then $IS(TTC) \dot{\leq} IS(DA)$ as for each pair $(i, o)$, $IS(TTC; i, o) \leq I(A_n^{TTC}) \leq I(\hat{A}_n^{DA}) \leq IS(DA; i, o)$.

For $n = 3$, we can apply Example 2 here, and thus we omit it.

## B.4 Proof of Proposition 2

The proof follows from the definitions of $IA^\triangleright$ and *favoring-higher-ranks*.

Let $f$ be a mechanism that satisfies *favoring-higher-ranks*. First, consider the case when all students' preferences are identical. In this case, to select the final output, there must be a tie-breaking rule over students (the rule may vary across different schools). This tie-breaking rule $\tau$ at least contains the informativeness of the $n-1$ schools, which is possibly related to the students' preferences. Let the informativeness information that we additionally have to know from this rule be $\zeta$. Clearly, the informational size from this rule will be minimized if it is fully depending on students' preferences (as we defined for $IA^\triangleright$). In other words, the information size of $IA^\triangleright$'s $\tau$ is smallest.

Next, consider $(i, o)$. If $o$ is $i$'s most preferred school, by *favoring-higher-ranks*, we know that $(i, o)$ is matched even in the worst case, if no other student positions $o$ at the top. Thus, except for $\zeta$, we additionally need to know all other students' first choice, $R^1_{-i}$. Similarly, if $o$ is $i$'s second most preferred school, then by *favoring-higher-ranks*, it is necessary to know all other students' first choice to confirm that $(i, o)$ is matched in the worst case, and all other students' (except for one who is matched with $i$'s first choice) second choice, and so on.

Overall, to confirm that $(i, o)$ is matched under $f$, for students' preferences we need to know $(n-1) + (n-2) + \ldots + 1$, plus the information about $\zeta$. Clearly, the sum of them will be at least as large as $IA^{\triangleright}$.

# Appendix C  General Framework

Here, we present a general framework to apply our measurement in a general environment.

## C.1  Preliminaries

There is a finite *set of agents* $N = \{1, \ldots, n\}$ and a *set of allocations* $X$. The generic elements in them are referred to $i \in N$ and $x \in X$, respectively. Note that an allocation might involve a monetary transfer. For each $i \in N$ and $x \in X$, let $x_i$ be $i$'s *allotment*, and $X_i$ be the set of $i$'s all possible allotments. For instance, in matching problems each agent's allotment is her matched partner, and in auction problems her allotment is her received objects and payment. For each allocation $x \in X$, we say that a subset of it is a *partial allocation* of $x$. Formally, $x'$ is a partial allocation of $x$ if for each $i \in N$, $x'_i \subseteq x_i$. For each $x \in X$, let $\mathscr{X}(x)$ be the set of partial allocations of $x$. In this way, we also recursively define the partial allocation's partial allocation: for any $y, z \in \mathscr{X}(x)$, $z$ is a partial allocation of $y$ if for each $i \in N$, $z_i \subseteq y_i$.

Each agent has a *type* that determines her evaluation of allocations. For instance, in matching with strictness, an agent's type is her ranking for another side and in auction with quasi-linearity her type is her valuation of objects. Let $T$ be the *set of types* and $T^N$ be the *set of type profiles*. Moreover, for each type, we construe a message that reveals partial information about it. For instance, in matching a message could be a partial ranking over objects, and in auction a message could be a range of valuations.[17] For each agent $i$ and her type $t_i \in T$, let $M(t_i)$ be the set of messages that reveals partial information about $t_i$, and let $\mathscr{M}$ be the *set of all possible messages*. Given a type profile $t \in T$, we refer to a message profile $m(t) \in \Pi_{i \in N} M(t_i)$ as a *partial type profile* of $t$ when there is no confusion. Conversely, for each partial type profile $m$, let $T^m = \{t \in T^N \mid m \in \Pi_{i \in N} M(t_i)\}$ be the set of type profiles for which $m$ can be viewed as the partial type profile of them. Recursively, a

---

[17]Consider the case where agent $i$ has the preference $R_i : a, b, c, d$ over objects $\{a, b, c, d\}$. In this case, $R_i^{a,c} : a, c$, i.e., the restriction of $R_i$ to $a, c$ is a message that indicates agent $i$ prefers $a$ to $c$. Similarly, in the case where agent $i$'s valuation of the item to be auctioned is $v_i \in \mathbb{R}_+$, the range $[0, 2 + v_i] \subsetneq \mathbb{R}_+$ is a message that reveals agent $i$'s valuation falls within this range.

partial type profile $m'$ is a partial type profile of $m$, if $T^m \subseteq T^{m'}$.[18]

An *algorithm* (allocation rule) selects an allocation for each type profile taking the profile as input and the selected allocation as output. Note that an algorithm may also output a partial allocation with some partial type profiles. A *mechanism* consists of two parts, a message space and an algorithm. Throughout the paper, we only focus on direct mechanisms. Thus, $\mathcal{M} = 2^T$ and the message space is $\mathcal{M}^N$. Therefore, since the message space remains fixed throughout, we will simply denote a mechanism $(\mathcal{M}^N, f)$ by an algorithm $f$.

## C.2 Informational Size

Now we introduce our measurement *informational size*, which represents the smallest piece of information that reveals agents' types and is required to implement a partial allocation in a mechanism. In other words, if we input information that is smaller than the associated *informational size*, the mechanism cannot output the desired partial allocation.

Given a mechanism $f$, let $t \in T^N$ and $y \in \mathcal{X}(f(t))$ be a partial allocation of $f(t) \in X$. We say that a partial type profile $m(t)$ is *sufficient to implement $y$ at $f$* if when we take $m(t)$ as input, $y$ is also a partial allocation of $f(m(t))$. We say that a partial type profile $m(t)$ is *necessary to implement $y$ at $f$* if for each partial type profile of $m(t)$, it is not sufficient to implement $y$ at $f$. Let $m^*(t)$ be a partial type profile that is both sufficient and necessary to implement $y$ at $f$ with the smallest cardinality.[19] We refer to the cardinality of $m^*(t)$ as the *informational size* to implement $y$ at $f$ that is with respect to $t$, denoted by $IS(f; y, t)$.

Next, we consider the informational size to implement $y$ at $f$ that is across all possible type profiles. Let $T^N(y) \subseteq T^N$ be the set of type profiles such that for each $\tau \in T^N(y)$, $y$ is a partial allocation of $f(\tau)$. Let $IS(f, y) = \max\{IS(f; y, \tau) \mid \tau \in T^N(y)\}$, i.e., it is the informational size in the worst-case scenario.

## C.3 Applications to Auction Problems

Consider a single item auction problem with two bidders. Now we apply our measurement of informational size to compare the simplicity of two static mechanisms (the first-price and the second-price auctions), and two dynamic mechanisms (the descending-clock and the ascending-clock auctions).

Here, we consider from the perspective of the winner, how she can verify the auction outcome. For bidder $i$, in order to confirm that she is the winner, $i$ only needs to know that she bids higher than her rival. The analysis reveals its subtlety when the winner attempts to confirm her payment. For first-price auction(FPA) and descending-clock auction (dA), since the payment is exactly her bidding, $i$ does not need any additional information. However, for second-price auction (SPA) and ascending-clock auction (aA), since the payment is the loser's bidding, $i$ has to know the exact value of the loser's bidding. It means that SPA (resp. aA) is *less informative* compared to FPA (resp. dA), as the latter needs more information.

---

[18]Alternatively, we can introduce nature as an agent in this general framework; for instance, suppose that agent 0 is the nature, $X_0 = \{\emptyset\}$ and $T_0$ is the set of states. This generalization allows us to treat the moving sequences in serial dictatorship mechanisms as the states drawn from nature.

[19]For the uncountable case, we define *smallest* by $\cap$, i.e., $m^*(t)$ is an union of all partial type profiles that are both sufficient and necessary to implement $y$ at $f$.

The comparison between FPA and SPA is not surprising, as it aligns with the existing literature, e.g., Li (2017) and Grigoryan and Möller (2023). However, the comparison between dA and aA is interesting: both are known to be credible in the sense of Akbarpour and Li (2020), however, our analysis highlights that since the seller still has the opportunity to overcharge under aA,[20] this possibility may raise concerns for buyers who are concerned about the worst-case scenario. Consequently, aA may be considered less credible than dA, as it is less transparent than dA in the sense that the winner needs to acquire additional information to confirm her payment.

---

[20]We would like to express our gratitude to Ruizhi Zhu for highlighting the following event. Suppose that the true payment is $p$ and the loser leaves when $p$ is offered. However, if the winner does not observe this, then the seller can still offer a higher price, say $p+\epsilon$, and the winner will take this offer since her valuation is higher than it. Note that when we consider the worst-case scenario, we only care whenever this event is possible, regardless of the expectation of this event. This is the reason why our finding differs from Akbarpour and Li (2020).